\documentclass[10pt,conference]{IEEEtran}
\IEEEoverridecommandlockouts
\usepackage{cite}
\usepackage{amsmath,amssymb,amsfonts}
\usepackage{algorithmic}
\usepackage{graphicx}
\usepackage{textcomp}
\usepackage{xcolor}
\usepackage{float}
\usepackage{subfig}
\usepackage{xspace}
\usepackage[linesnumbered,ruled,boxed]{algorithm2e}
\usepackage{enumitem}
\usepackage{siunitx}
\usepackage{array}
\usepackage{booktabs}
\usepackage[hyphens]{url}
\usepackage{hyperref}
\usepackage{balance}

\graphicspath{{figs/}}

\newcolumntype{H}{>{\setbox0=\hbox\bgroup}c<{\egroup}@{}} 

\newcommand{\ours}{EALink\xspace}
\newcommand{\oursf}{EALink$_f$\xspace}

\DeclareRobustCommand{\IEEEauthorrefmark}[1]{\smash{\textsuperscript{\footnotesize #1}}}
\pdfoutput=1
\begin{document}

\title{EALink: An Efficient and Accurate Pre-trained Framework for Issue-Commit Link Recovery}

\author{
\IEEEauthorblockN{
Chenyuan Zhang\IEEEauthorrefmark{1},
Yanlin Wang\IEEEauthorrefmark{2}, 
Zhao Wei\IEEEauthorrefmark{3},
Yong Xu\IEEEauthorrefmark{3},
Juhong Wang\IEEEauthorrefmark{3},
Hui Li\IEEEauthorrefmark{1}\textsuperscript{*} and
Rongrong Ji\IEEEauthorrefmark{1}
}
\IEEEauthorblockA{
\IEEEauthorrefmark{1}\textit{Key Laboratory of Multimedia Trusted Perception and Efficient Computing, Ministry of Education of China}\\
\textit{School of Informatics, Xiamen University, China}
}
\IEEEauthorblockA{
\IEEEauthorrefmark{2}\textit{School of Software Engineering, Sun Yat-sen University, China}
}
\IEEEauthorblockA{
\IEEEauthorrefmark{3}\textit{Tencent, China}
}
\IEEEauthorblockA{
zhangchenyuan@stu.xmu.edu.cn,\,\,\,\,wangylin36@mail.sysu.edu.cn,\\
\{zachwei,\,rogerxu,\,julietwang\}@tencent.com,\,\,\,\,\{hui,\,rrji\}@xmu.edu.cn
}
}


\maketitle

\begingroup\renewcommand\thefootnote{*}

\footnotetext{Corresponding author.}
\endgroup

\begin{abstract}
	Issue-commit links, as a type of software traceability links, play a vital role in various software development and maintenance tasks. However, they are typically deficient, as developers often forget or fail to create tags when making commits. Existing studies have deployed deep learning techniques, including pre-trained models, to improve automatic issue-commit link recovery. Despite their promising performance, we argue that previous approaches have four main problems, hindering them from recovering links in large software projects. To overcome these problems, we propose an efficient and accurate pre-trained framework called EALink for issue-commit link recovery. EALink requires much fewer model parameters than existing pre-trained methods, bringing efficient training and recovery. Moreover, we design various techniques to improve the recovery accuracy of EALink. We construct a large-scale dataset and conduct extensive experiments to demonstrate the power of EALink. Results show that EALink outperforms the state-of-the-art methods by a large margin (15.23\%-408.65\%) on various evaluation metrics. Meanwhile, its training and inference overhead is orders of magnitude lower than existing methods. We provide our implementation and data at https://github.com/KDEGroup/EALink.
\end{abstract}



\section{Introduction}
\label{sec:intro}

Software traceability links maintain associations between diverse artifacts (e.g., requirements, design and source code)~\cite{AntoniolCCLM02} and support various software development and maintenance tasks, e.g., software change impact analysis~\cite{AungHS20}, identifying vulnerability-fixing commits~\cite{Nguyen-TruongKL22}, selective regression testing~\cite{NaslavskyR07} and project management~\cite{Panis10}.
Unfortunately, the maintenance of software traceability links typically relies on the labor-intensive manual work of developers, leading to incomplete, inaccurate and even conflicting traceability links~\cite{RodriguezCF21}.
Hence, the deficient software traceability links increase the cost of software development and maintenance.

\begin{figure}[t]
\centering
\includegraphics[width=0.98\columnwidth]{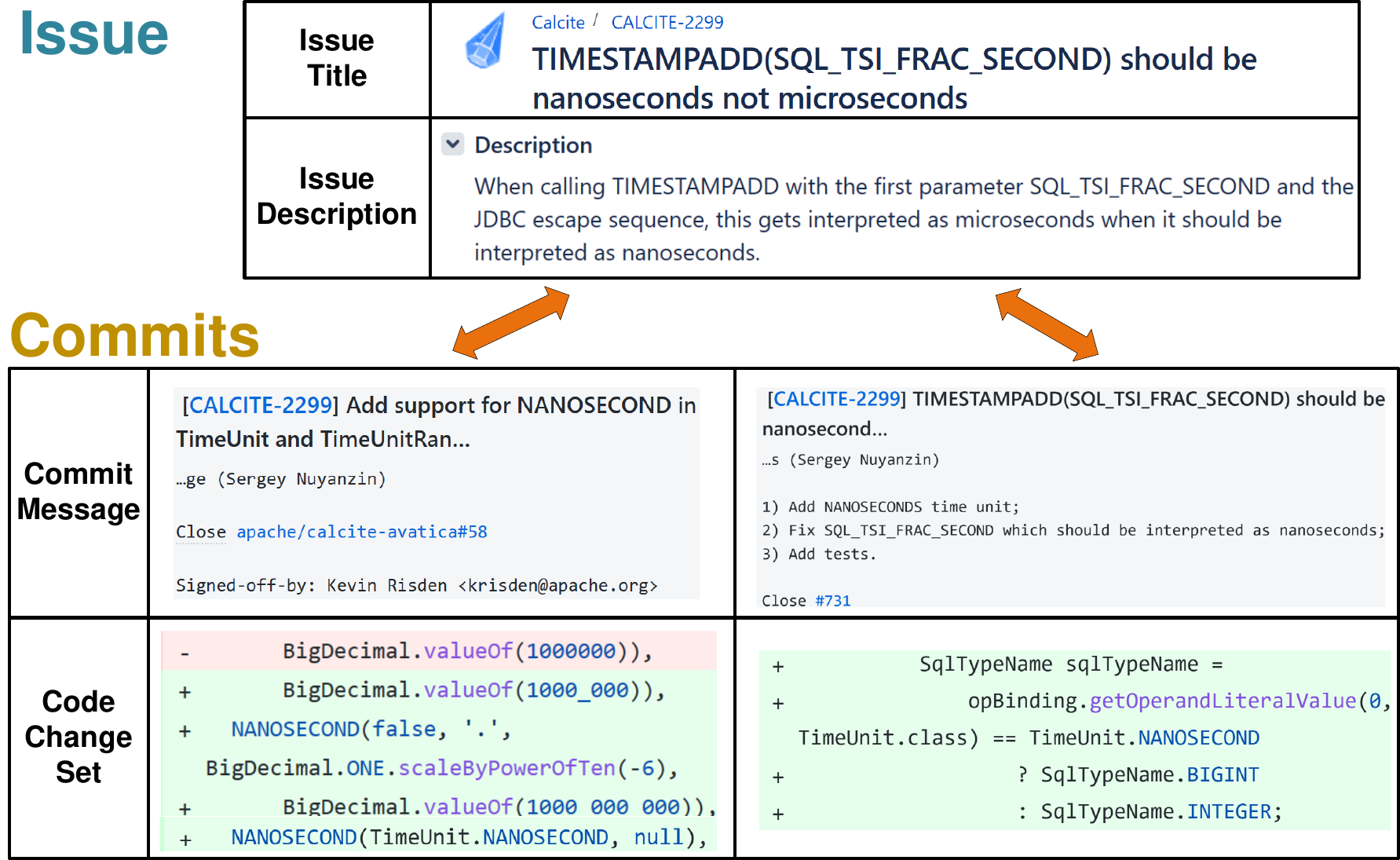}
\caption{An example of an one-to-many issue-commit link.}
\label{fig:iclink_example}
\end{figure}

In this paper, we focus on recovering software traceability links among related issues and commits in software repositories, i.e., \emph{issue-commit link recovery}.
Issues summarize user's discussions around required changes of the software in the form of documentation, while commits contain the change itself with a commit message using natural language text~\cite{MazraeIH21}.
With the ubiquitous adoption of version control systems and platforms such as Git and GitHub, and issue tracking systems such as Jira and Bugzilla, 
developers can tag commits with corresponding issues as they perform daily software development and maintenance tasks~\cite{0002RGCM18}. 
An issue is commonly fixed by one or more commits~\cite{SunCWB17}.
Fig.~\ref{fig:iclink_example} shows the data format of an issue-commit link with two commits are linked to the issue via the issue tag [CALCITE-2299], we explain details of the data collection process in Sec.~\ref{sec:data}.
The traceability links between issues and commits play a vital role in software development and maintenance activities (e.g., commit analysis~\cite{RuanCPZ19} and bug prediction~\cite{WuZKC11}).
Nevertheless, issue-commit links suffer from the same issue as other software traceability links: links are typically deficient as developers often forget or fail to create tags when making commits~\cite{abs-2211-00381}.

\begin{figure*}[t]
\centering
\includegraphics[width=0.98\linewidth]{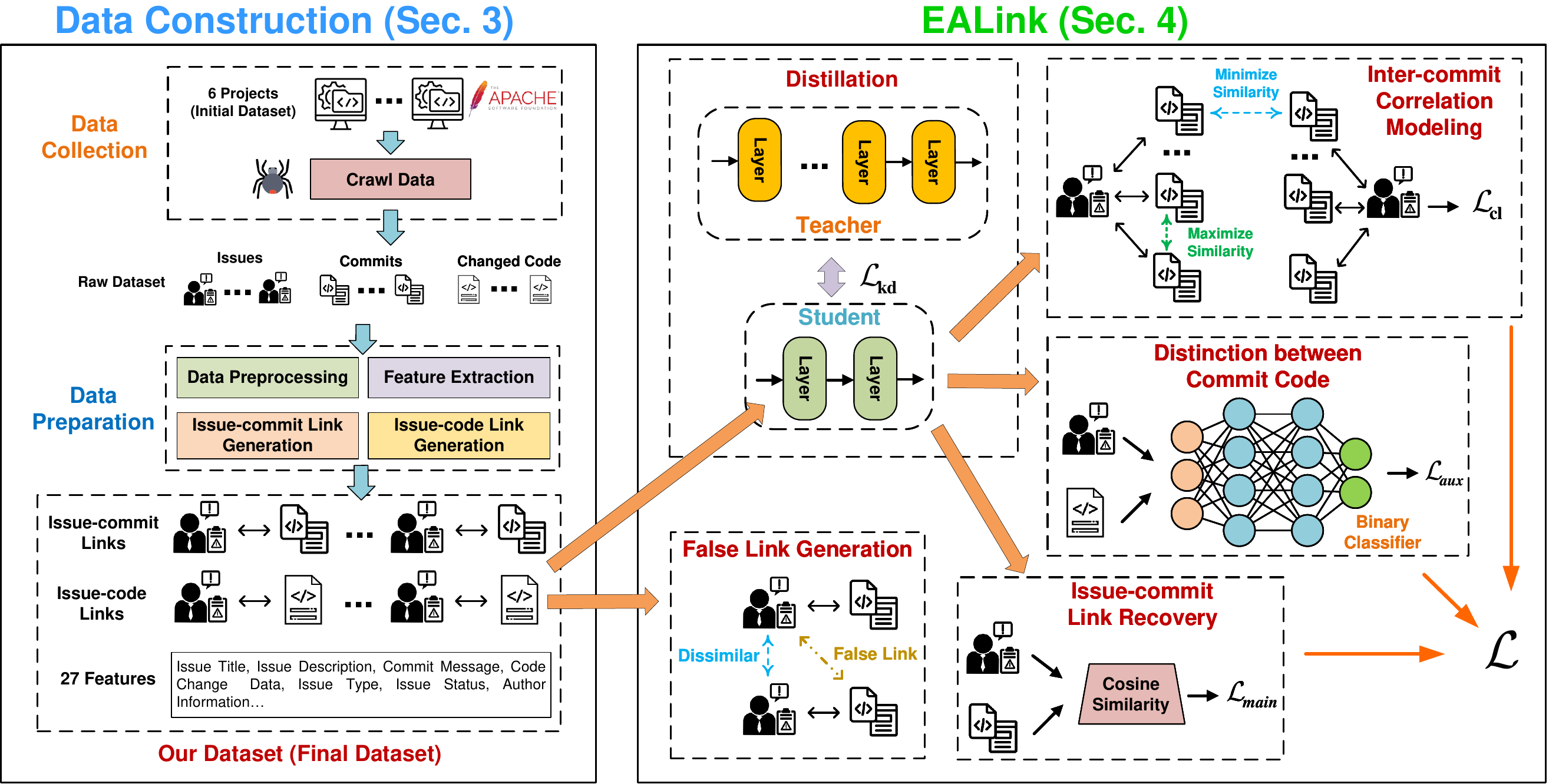}
\caption{Overview of this work. Orange arrows indicate the relationship among different parts in \ours.}
\label{fig:framework}
\end{figure*}

To alleviate the scarcity of issue-commit links, various automatic issue-commit link recovery methods are proposed.
Early approaches adopt feature-based and rule-based methods~\cite{WuZKC11,NguyenNNN12,SchermannBPLG15} which heavily rely on feature engineering and manual rules.
They show low precision and are hard to generalize.
Traditional learning based approaches~\cite{LeVLP15,SunWY17,SunCWB17,0002RGCM18,MazraeIH21} leverage machine learning to automatically learn from features and alleviate the reliance on manual rules.
More recently, researchers are inspired by the success of deep learning in various applications and adopt deep neural networks to enhance issue-commit link recovery~\cite{RuanCPZ19,XieCYLHDZ19,LinLZ0C21}. 
Deep learning methods encode code and textual data into separate spaces for modeling programming language (PL) and natural language (NL) in order to overcome the semantic gap~\cite{LuciaFO08,0004CC17} between different software artifacts.
Moreover, they can leverage pre-training techniques and pre-train the model on related software engineering tasks and code corpora so that the issue-commit link recovery task can benefit from large-scale code data and supervision beyond issue-commit link data~\cite{LinLZ0C21}.

Although much effort has been devoted to interlinking issues and commits, we find that there still exist problems:
\begin{enumerate}
	\item \textbf{P1: High Training and Inference Overhead:} Despite the success of deploying pre-trained models (e.g., BERT~\cite{DevlinCLT19}) in issue-commit recovery~\cite{LinLZ0C21}, the performance improvement comes at the cost of much more model parameters and longer training and inference time, hindering their use in large-scale software projects. For example, T-BERT~\cite{LinLZ0C21} takes 44,353 seconds for 1,000 recovery queries on a moderate-size dataset in our experiments.

	\item \textbf{P2: Neglect of Inter-commit Correlation:} In practice, multiple commits may correspond to the same issue. Prior methods mostly model links between an issue and each corresponding commit, neglecting correlations among commits in a one-to-many issue-commit link (i.e., inter-commit correlation).

	\item \textbf{P3: Nondistinctive Modeling of Changed Code in One Commit:} Existing approaches treat all changed code files in a commit equally. However, this is not reasonable in practice. Analysis on code repositories demonstrates that changes for different purposes can be submitted together in a single commit~\cite{SunWY17}. Inappropriately modeling loosely related and unrelated code changes introduces noise, jeopardizing learning quality.

	\item \textbf{P4: Conflicting False Links:} Many deep learning based recovery approaches are trained via separating true links and false links. True links that correctly connect issues to their related commits can be extracted from tags provided on GitHub, Jira and Bugzilla.
	To construct inexistent false links that interlink issues and unrelated commits, existing methods adopt a sampling method based on the time interval.
	However, such a method may generate false links that are actually true links. See Sec.~\ref{sec:motivation} for details.

\end{enumerate}

We provide motivating examples for each problem in Sec.~\ref{sec:motivation}. 
To tackle the above problems, we propose an \underline{E}fficient and \underline{A}ccurate Pre-trained framework for issue-commit \underline{Link} recovery (\ours). 
Fig.~\ref{fig:framework} provides an overview of this work and our contributions are:
\begin{itemize}

\item To reduce the overhead brought by large model size (\textbf{P1}), \ours distills knowledge from a pre-trained NL-PL model (e.g., CodeBERT~\cite{FengGTDFGS0LJZ20}) to construct a compact model.
The distilled model can capture the semantic connections between NL and PL, which is essential to model issues and commits.
Meanwhile, it is easier to fine-tune the compact model on the issue-link recovery task since it contains fewer parameters and requires much shorter training and inference time.

\item To model inter-commit correlation (\textbf{P2}), \ours employs contrastive learning~\cite{LiuZHMWZT23}. And inter-commit correlation is captured by contrasting positive commits (from the same one-to-many issue-commit link) and negative commits (from different issue-commit links).

\item To provide a fine-grained distinction between data of changed code files (\textbf{P3}), we design an auxiliary task \emph{issue-code link prediction} to help distinguish the importance of each commit code.
The task is jointly trained with the issue-commit link recovery task in a manner of multi-task learning.

\item To avoid conflicting false links (\textbf{P4}), we propose a false link generation mechanism for constructing reasonable false links used for training \ours.

\item To fairly evaluate \ours and existing works on large-size projects, we construct a new, large dataset for the issue-commit link recovery task (it also contains issue-code links for our designed auxiliary task). Extensive experiments demonstrate that \ours recovers issue-commit links more efficiently and accurately than state-of-the-art baselines on large projects.

\end{itemize}


\section{Motivating Examples}
\label{sec:motivation}

In this section, we provide detailed motivating examples to illustrate the four problems mentioned in Sec.~\ref{sec:intro}.

\vspace{5pt}
\noindent\textbf{Example for P1 (High Training and Inference Overhead):} 
Recent deep learning based issue-link recovery models show promising performance~\cite{RuanCPZ19,LinLZ0C21} as they can better capture textual features, code features and their mutual relations via deep neural networks.
However, the strong expressive power of deep neural networks, especially pre-trained models~\cite{DevlinCLT19, abs-1907-11692}, comes at the cost of requiring many more model parameters and longer training/inference time than traditional methods. 
For instance, deep learning based method DeepLink~\cite{RuanCPZ19} and pre-training based method T-BERT~\cite{LinLZ0C21} take 1,145 seconds and 44,353 seconds for inference on Isis, a moderate-size project in our constructed dataset. And T-BERT requires about 138 hours for pre-training using our hardware environment.
The high training and inference cost of deep learning based recovery methods limits their use in processing large software projects.

\begin{figure}[t]
	\centering
	\includegraphics[width=0.95\columnwidth]{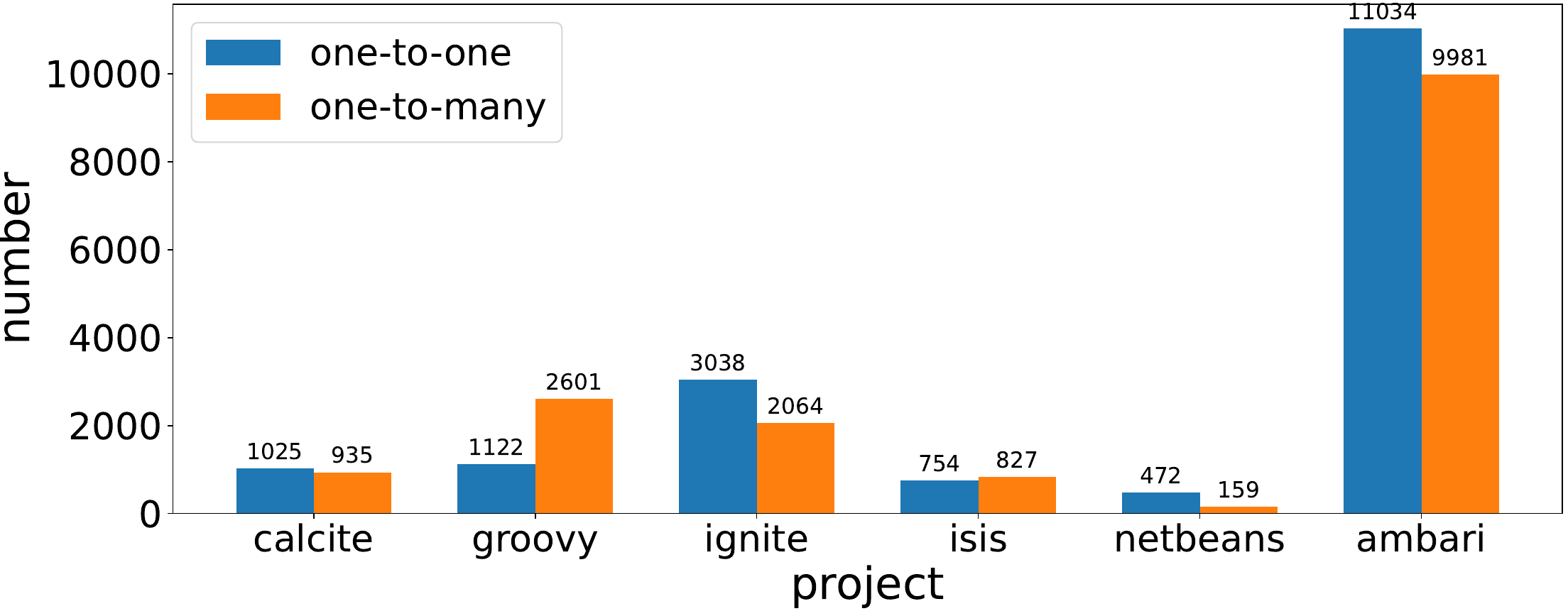}
	\caption{Statistics of one-to-one and one-to-many links.}
	\label{fig:one-to-many_dis}
\end{figure}

\vspace{5pt}
\noindent\textbf{Example for P2 (Neglect of Inter-Commit Correlation):}
In a software project, developers may make multiple commits for fixing an issue, forming one-to-many issue-commit links.
One-to-many links are ubiquitous.
Fig.~\ref{fig:one-to-many_dis} demonstrates the statistics of one-to-one and one-to-many links in the 6 projects in our constructed data (Sec.~\ref{sec:data}). 
We can observe that all the 6 software projects have roughly similar numbers of one-to-one and one-to-many links.
Multiple commits in a one-to-many link may be correlative.
In the example of Fig.~\ref{fig:iclink_example}, the two commits together fix the issue [CALCITE-2299].
However, existing works will model the two commits separately, ignoring that they are related to each other.

\begin{figure}[t]
\centering
\includegraphics[width=0.95\columnwidth]{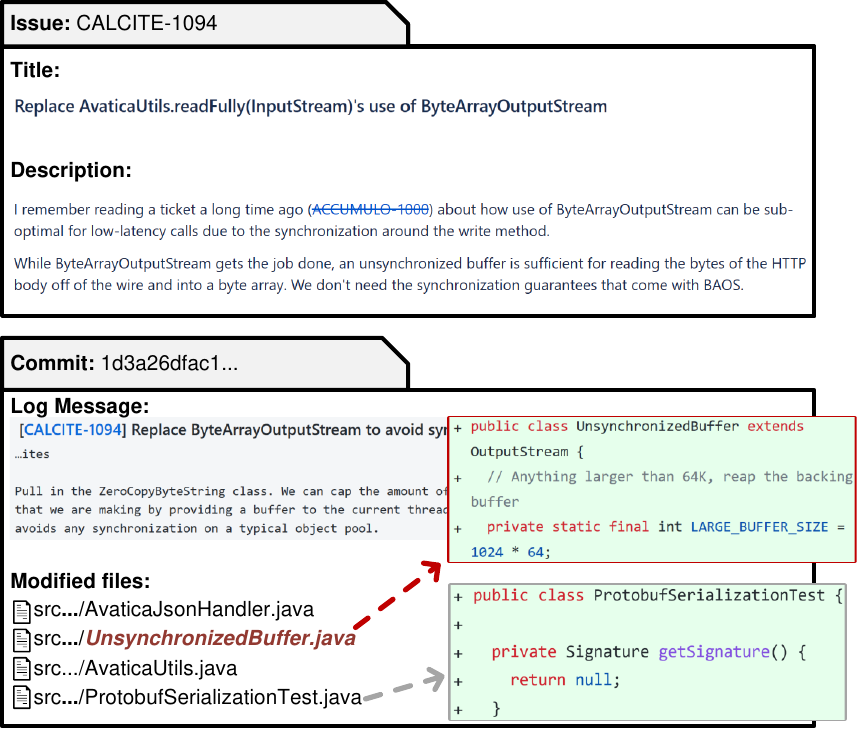}
\caption{Loosely related or unrelated code in a commit.}
\label{fig:diff_code}
\end{figure}

\vspace{5pt}
\noindent\textbf{Example for P3 (Nondistinctive Modeling of Changed Code in One Commit):} 
Commits often contain changed code located in different source code files and some of them are loosely related or even unrelated to the issue. 
Fig.~\ref{fig:diff_code} depicts a commit for the issue [CALCITE-1094].
The changes of the source code file ``UnsynchronizedBuffer.java'' is relevant to [CALCITE-1094], while the changes of ``ProtobufSerializationTest.java'' implement a test program and they are loosely related or unrelated to [CALCITE-1094].
Loosely related or unrelated source code files introduce noise to issue-commit link recovery.
Noise hinders recovery model from capturing critical information.
Most existing approaches do not take special actions to handle loosely related or unrelated source code file data.
A few works~\cite{SunWY17} filter changed code that does not share many terms with issues to avoid noise.
But their methods rely on keyword matching that is inappropriate for solving this problem:
a) Commits (code) and issues (title and description) are in different modalities (NL and PL) that do not share the same vocabulary;
b) Commits and issues may contain relevant but not exactly matching terms and using keyword matching will filter such useful, related code snippets.

\vspace{5pt}
\noindent\textbf{Example for P4 (Conflicting False Links):}
Previous works generate false links by sampling commits submitted 7 days before or after one of the three dates (creation/updated/resolved dates) of an issue and connect the sampled commits to the issue as false links~\cite{BirdBRB10,NguyenNNN12,SunWY17,RuanCPZ19}.
However, in practice, many factors including active developers and simple issues make it possible that some commits are submitted shortly after the creation of the issue.
For instance, the issue [CALCITE-1700] has a commit\footnote{\raggedright \url{https://github.com/apache/calcite/commit/8bac70e5020c50505b34df6eaae18c0ca414f2c2}} submitted within 7 days after the creation of [CALCITE-1700].
Therefore, unlabeled true links may be wrongly sampled as false links using time interval based sampling method adopted by previous works.


\section{Dataset Construction}
\label{sec:data}

Existing works adopt small datasets for evaluation.
For example, Lin et al.~\cite{LinLZ0C21} uses three software projects with hundreds of issues, commits and links in the evaluation.
However, the effectiveness and efficiency of recovery models cannot be fully evaluated on small-scale data.

Hence, we constructed a large issue-commit link dataset based on 6 Java projects to evaluate link recovery methods.
The data for each project contains true links ranging from one thousand three hundred to thirty five thousand.
Moreover, it includes 27 non-text and text features.
Tab.~\ref{tab:dataset} depicts our constructed final dataset and Tab.~\ref{tab:feature} provides detailed descriptions of the 27 features.
Note that not all the features are used in this study. 
Unused features could be helpful for future studies on issue-commit link recovery.

This section shows the steps of dataset construction which are depicted in the left of Fig.~\ref{fig:framework}.
For the purpose of illustration, in the following sections, \emph{a link refers to a one-to-one issue-commit link and a one-to-many issue-commit link is broken into multiple one-to-one issue-commit links.}

\begin{table}[t]
\caption{Statistics of our constructed final dataset.}
\label{tab:dataset}
\centering
\begin{tabular}{@{}ccccc@{}}
\toprule
\textbf{Project} & \textbf{\#Issues} & \textbf{\#Commits} & \textbf{\begin{tabular}[c]{@{}c@{}}\#True Issue-\\ commit Links\end{tabular}} & \textbf{\begin{tabular}[c]{@{}c@{}}\#True Issue-\\ code Links\end{tabular}} \\ \midrule
Ambari           & 25,162            & 38,872             & 35,597                                                                        & 52,530                                                                      \\
Calcite          & 3,740             & 6,934              & 3,058                                                                         & 7,910                                                                       \\
Groovy           & 9,118             & 30,633             & 8,851                                                                         & 9,208                                                                       \\
Ignite           & 12,495            & 32,930             & 9,997                                                                         & 17,747                                                                      \\
Isis             & 2,264             & 15,284             & 8,486                                                                         & 27,127                                                                      \\
Netbeans         & 3,705             & 19,181             & 1,369                                                                         & 4,662                                                                       \\ \bottomrule
\end{tabular}
\end{table}

\begin{table}[t]
\centering
\caption{Descriptions of 27 features in final dataset.}
\label{tab:feature}
\begin{tabular}{|c|l|}
\hline
\textbf{Feature}   & \multicolumn{1}{c|}{\textbf{Description}}                                                                             \\ \hline
source             & \begin{tabular}[c]{@{}l@{}}Describe which platform or system the \\ project originates from (e.g., apache).\end{tabular} \\ \hline
product            & Project name (e.g. Netbeans).                                                                                           \\ \hline
issue\_id           & \begin{tabular}[c]{@{}l@{}}ID number corresponding to an issue \\ (e.g., 13226408).\end{tabular}                         \\ \hline
component          & \begin{tabular}[c]{@{}l@{}}The component type that the project belongs \\ to (e.g., engine).\end{tabular}                     \\ \hline
creator\_key        & \begin{tabular}[c]{@{}l@{}}A hash value that uniquely identifies \\ the issue (e.g., 7860ba1e91f42c...).\end{tabular}    \\ \hline
create\_date        & \begin{tabular}[c]{@{}l@{}}The creation time of the issue \\ (e.g., 2018-03-13 09:06:55+00:00).\end{tabular}                     \\ \hline
update\_date        & The last update time of the issue.                                                                                    \\ \hline
last\_resolved\_date & The last resolved time of the issue.                                                                                  \\ \hline
comment            & Discussion on the issue.                                                                                         \\ \hline
summary            & Summary of the issue.                                                                                                     \\ \hline
description        & The specific content of the issue.                                                                                    \\ \hline
issue\_type        & \begin{tabular}[c]{@{}l@{}}The type of the issue \\ (e.g., bug/improvement/new feature).\end{tabular}                        \\ \hline
status             & \begin{tabular}[c]{@{}l@{}}The status of the issue \\ (e.g., open/close/resolved).\end{tabular}                            \\ \hline
repo               & \begin{tabular}[c]{@{}l@{}}The name of the project that the commit \\ belongs to.\end{tabular}                       \\ \hline
commitid           & \begin{tabular}[c]{@{}l@{}}A hash value that uniquely identifies  \\ the commit. \end{tabular}                          \\ \hline
parents            & The hash value of the lasted submission.                               \\ \hline
author             & Hash for the author of commit.                                                                                        \\ \hline
committer          & \begin{tabular}[c]{@{}l@{}}The hash of the person who made   \\ the commit. \end{tabular}                                \\ \hline
author\_time\_date   & \begin{tabular}[c]{@{}l@{}}The time when the author submitted   \\ the commit. \end{tabular}                           \\ \hline
commit\_time\_date   & \begin{tabular}[c]{@{}l@{}}The time when the committer submitted \\ the commit.\end{tabular}                            \\ \hline
message            & Summary of the content of the commit.                                                                                  \\ \hline
commit\_issue\_id    & \begin{tabular}[c]{@{}l@{}}If the commit has been marked, the  \\ value of commit\_issue\_id is the issue id.\end{tabular}                      \\ \hline
changed\_files      & The path of the changed file.                                                                                         \\ \hline
Diff               & Code diff.                                                                                          \\ \hline
codelist           & List of source code files.                                                                                            \\ \hline
nonsource          & Content of changed non-source code files.                                                                                      \\ \hline
label              & \begin{tabular}[c]{@{}l@{}}Its value is 1 if it corresponds to a true \\  link, otherwise 0.\end{tabular}                                                                                  \\ \hline
\end{tabular}
\end{table}

\subsection{Data Collection}

Claes et al.~\cite{ClaesM20} present a dataset containing common issue and commit information (i.e., issue title, issue descriptions, issue comments, commit messages and meta-information of changed source code files) on 765 projects across 20 years. 
However, it lacks information of code change data and source code file data in commits. 
We selected 6 Apache projects from their dataset based on project popularity (i.e., the number of stars and forks) as the \emph{initial dataset}.
After that, we crawled code change data and source code data based on the meta-information of changed source code files that contains urls of code data.
The crawled code data was added to the initial dataset as a supplement to form the \emph{raw dataset}.

\subsection{Data Preparation}

We prepared the \emph{final dataset} by conducting four operations on the raw dataset: data preprocessing, feature extraction, true issue-commit link generation and issue-code link generation.

\subsubsection{Data Preprocessing}
The raw dataset contains information (e.g, issue description) created by different developers with diverse development conventions.
Hence, different text styles and programming styles are manifested in the raw dataset, bringing noise.
To improve data quality, we adopted several strategies to preprocess textual and code data contained in the raw dataset.

\vspace{5pt}
\noindent\textbf{Text Preprocessing:}
The textual data in the raw dataset contains issue title, issue description and commit message. 
We first performed several common NLP preprocessing strategies including lowercase conversion, tokenization, stop word removal and stemming~\cite{Palmer10} on textual data.
They not only reduced the size of the token vocabulary, thus allowing for a compact feature set, but also integrated different forms of words by replacing them with root words. 
Then, following DeepLink~\cite{RuanCPZ19}, we removed hyperlinks and issue tags from textual data.
Hyperlinks were removed since they are typically not viewed as textual data.
Issue tags were removed since all commit messages in some project repositories\footnote{The Calcite repository is one example: \url{https://github.com/apache/calcite}.} contain issue tags (e.g., the commit message in Fig.~\ref{fig:iclink_example}). Since issue tags can be directly used to find true links (see Sec.~\ref{sec:true_link_gen}), we removed them to avoid data leakage and increase the difficulty of issue-commit link recovery.
We also identified inline code and large code blocks from textual data using regular expressions.
Identified issue code was removed from textual data but was included as features in the final dataset (see Sec.~\ref{sec:feat_extract}).

\vspace{5pt}
\noindent\textbf{Code Preprocessing:} 
We first extracted identifier names (e.g., class names, method names and variable names) from code data. 
Moreover, for source code data, since it contains the complete function body,
we used tree-sitter\footnote{\url{https://github.com/tree-sitter/tree-sitter}} to convert it to Abstract Syntax Tree (AST) and extract identifier names from node tokens. 
For code change data, we used the patterns (i.e., regular expressions) proposed by Nguyen et al.~\cite{NguyenNNN12} and Sun et al.~\cite{SunWY17} to extract identifier names. 
Then, the extracted identifier names were split into tokens according to their patterns.
Finally, all code data was converted to lowercase.

\subsubsection{Feature Extraction}
\label{sec:feat_extract}
The features extracted and used by \ours are \emph{issue title}, \emph{issue description}, \emph{commit message}, \emph{code change} and \emph{changed source code files in each commit}.
Note that this paper mainly focuses on the model design for the link-commit issue recovery task and \ours only considers the above five features.
We additionally extracted many other textual and non-textual features (e.g., issue status and updated date) and included them in the final dataset since some issue-commit link recovery approaches require additional features.
For instance, DeepLink~\cite{RuanCPZ19} requires \emph{issue code} which includes code tokens in issue descriptions.
KG-DeepLink~\cite{XieCYLHDZ19} uses a feature vector including issue type, issue priority, issue reporter, issue assignee, issue created time, issue resolved time, commit time and committer.
In total, our final dataset contains 27 features and these features can be useful to future study of link-commit issue recovery.

\subsubsection{True Issue-Commit Link Generation}
\label{sec:true_link_gen}
If a commit is marked with an issue tag, we labeled it together with the corresponding issue as a \emph{true link} and added to the final dataset. 
False links were generated during training (see Sec.~\ref{sec:opt}).

\subsubsection{Issue-Code Link Generation}
\label{sec:auxiliary_data}

\ours is also trained on the auxiliary task issue-code link prediction which will be described in Sec.~\ref{sec:mtl}.
To prepare the data for the auxiliary task, for each true issue-commit link, we simply checked whether the file name of each changed source code file appeared in issue title or issue description.
If so, we constructed an issue-code link between the changed source code file and the corresponding issue, and added it to the final dataset.

\section{Our framework \ours}
\label{sec:method}

\subsection{Overview}
The right part of Fig.~\ref{fig:framework} depicts \ours which contains two steps. 
In the first step, \ours distills knowledge of a large teacher model to construct a compact student model (Sec.~\ref{sec:kd}).
In the second step, \ours fine-tunes the lightweight student model for the task of issue-commit link recovery.
Specifically, during fine-tuning, \ours is trained to capture inter-commit correlation via contrastive learning (Sec.~\ref{sec:cl}), and distinguish the varying importance of different commit code through the auxiliary task of issue-code link prediction (Sec.~\ref{sec:mtl}).
In fine-tuning, \ours is optimized in a manner of multi-task learning and we propose a false link generation approach to prepare the links for training (Sec.~\ref{sec:opt}).

\vspace{5pt}
\noindent\textbf{Model Input:} The input to the first step (distillation) is a code pre-trained model. The input to the second step (fine-tuning) includes \emph{issue text} (i.e., the concatenation of the title and description of the issue), \emph{commit message}, \emph{commit code} (i.e., code data in one changed file of the commit) and \emph{code change} (i.e., difference between two versions of the changed code file).

\vspace{5pt}
\noindent\textbf{Notation:} 
$\mathcal{B}=\{t_1,t_2,\cdots,t_{\left| \mathcal{B} \right|} \}$ indicates a batch of issue-commit links, where $t_i=\{s_i, q_i\}$ and $t_i \in \mathcal{B}$.
$s_i$ and $q_i$ denote the issue and the commit in the one-to-one issue-commit link $t_i$.
$\mathcal{B}$ includes both true links $\mathcal{B}_\text{true}$ and false links $\mathcal{B}_\text{false}$.
The complete input data $X$ is randomly divided into multiple batches (i.e., multiple $\mathcal{B}$) in each training iteration following the standard machine learning training paradigm.
Other notations are explained when they are used.

\subsection{Distillation of Code Pre-training Model (P1)}
\label{sec:kd}

The software engineering community has embraced the use of pre-trained models that have achieved a great success in Natural Language Processing and Computer Vision domains~\cite{DevlinCLT19,ChenZHCSXX23}, and there is a surge of works on pre-trained models~\cite{FengGTDFGS0LJZ20,GuoRLFT0ZDSFTDC21,GuoLDW0022} for code-related tasks recently. 
For the issue-commit link recovery task, pre-training has been firstly adopted in T-BERT~\cite{LinLZ0C21}.

The effectiveness of pre-trained models can be attributed to their ability to learn universal representations from a massive amount of mostly unlabeled data through self-supervised learning tasks to benefit various downstream tasks with relatively much less data. 
This leads to a better starting point for model optimization, improved generalization performance, and a form of regularization to prevent overfitting on small datasets~\cite{abs-2003-08271}.
Despite the empirical success, a widely recognized problem of pre-trained models is the computational efficiency problem, as they often have a large number of parameters and take a long time for training and inference~\cite{SunCGL19,JiaoYSJCL0L20}.

\begin{figure}[t]
\centering
\includegraphics[width=0.7\columnwidth]{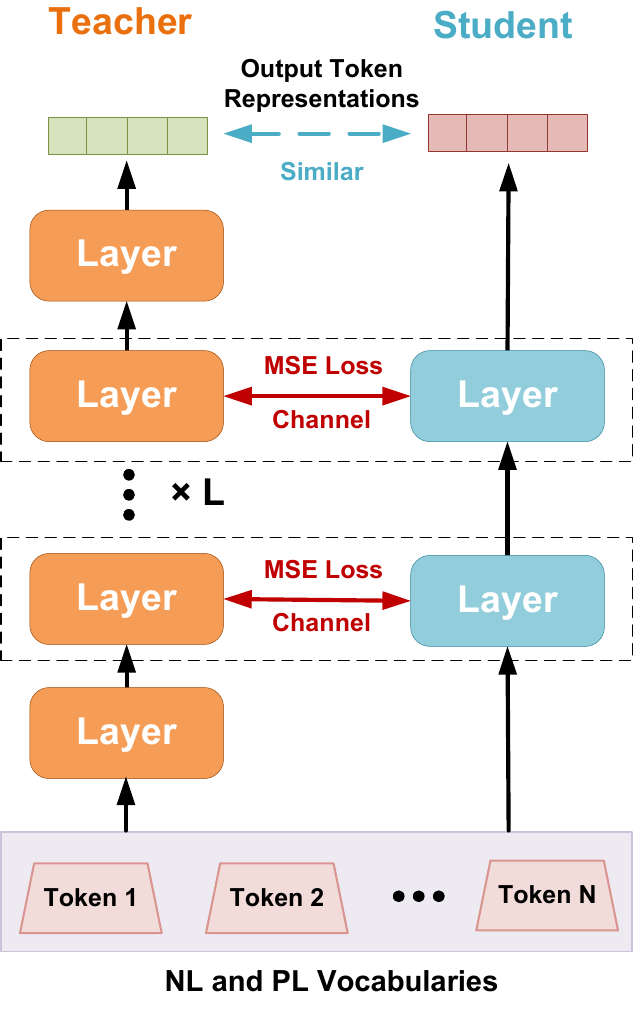}
\caption{Distillation for the code pre-trained model.}
\label{fig:kd}
\end{figure}

\ours is also designed based on pre-training and it can adopt any code pre-trained model to have all the benefits of pre-training.
To reduce the size of the pre-trained model as well as the training and inference time, we adopt the idea of knowledge distillation~\cite{GouYMT21} to transfer the knowledge encoded in the large pre-trained model (i.e., teacher) to a compact student model.
The student model is later used for issue-commit link recovery in \ours.
Fig.~\ref{fig:kd} illustrates the distillation process.

\vspace{5pt}
\noindent\textbf{Teacher Model:} Let $f_t(X;\theta)$ denote the code pre-trained teacher model where $X$ is the input to \ours and $\theta$ is model parameters.
Prevalent code pre-trained models typically adopt a BERT-style architecture which contains a multi-layer bidirectional Transformer~\cite{VaswaniSPUJGKP17} with $L$ Transformer blocks.
For instance, CodeBERT~\cite{FengGTDFGS0LJZ20} uses exactly the same architecture as RoBERTa-base~\cite{abs-1907-11692} with 12 layers.
Since the use of Transformer has become ubiquitous, we omit the background description of Transformer.

\vspace{5pt}
\noindent\textbf{Student Model:} The distillation component aims to learn a much smaller parameter set $\theta'$ for a compact student model $f_s(X;\theta')$ which shows similar performance to the large pre-trained teacher model.
In \ours, we use a two-layer RoBERTa (i.e., two Transformer blocks) as the student model.

\vspace{5pt}
\noindent\textbf{Distillation Strategy:}
A common and easy-to-deploy distillation strategy is the last-layer distillation~\cite{GouYMT21}, i.e., the student imitates the output from the last layer of the teacher model.
However, as the number of training epochs increases, the student model using last-layer distillation may face the poor generalization issue caused by overfitting on the training data~\cite{SunCGL19}.
Therefore, we opt to the intermediate-layer distillation strategy~\cite{RomeroBKCGB14} that distills encoded knowledge from intermediate layers in the teacher model.
The hidden states of all NL and PL tokens in a teacher's hidden layer are used as hints to train the  hidden states in a student's layer so that the hidden states of corresponding teacher and student layers are close.
To avoid high overhead, we train each of the two layers in the student model to mimic one intermediate layer in the teacher model and each teacher-student distillation layer pair is called as a distillation \emph{channel} as shown in Fig.~\ref{fig:kd}.
We use a dimension-wise MSE loss (i.e., squared Euclidean distance) in distillation:
\begin{equation}
	\label{eq:inter_kd}
	\mathcal{L}_{\text{kd}}=\sum_{\left<l_t,l_s \right>\in C} \sum_{v\in \mathcal{V}} \sum_{z=1}^d \left(\mathbf{h}^{(l_t)}_{v,z} - \mathbf{h}^{(l_s)}_{v,z}\right)^2,
\end{equation}
where $C$ denotes all pre-set distillation channels, $\mathcal{V}$ is the vocabulary containing all pre-trained NL and PL tokens, and $d$ indicates the dimensionality of hidden states. 
$\mathbf{h}^{(l_t)}_{v,z}$ and $\mathbf{h}^{(l_s)}_{v,z}$ indicates the $z$-th dimension of the hidden state for the token $v$ output in the $l_t$-th layer of the teacher model and the $l_s$-th layer of the student model, respectively. 
After the distillation, the student model can produce representations for NL and PL tokens that are similar to the teacher model's output representations.
And the student model has much fewer parameters, making it more easy to adopt the student model in fine-tuning and inference.

\subsection{Inter-Commit Correlation Modeling (P2)}
\label{sec:cl}

As discussed in previous sections, multiple commits may correspond to the same issue and we believe this provides additional supervision to help better model commits.
Since commits belonging to the same one-to-many issue-commit link should be related to each other (e.g., they together fix a bug), we design an inter-commit correlation modeling component in \ours to capture such correlations.

Specifically, this component is designed with the idea of contrastive learning~\cite{LiuZHMWZT23}.
Contrastive learning has become prevalent representation learning paradigm as it does not require labels and can learn such a representation space where similar sample pairs stay close to each other while dissimilar ones are far apart. 
For a commit $q_i$ in an issue-commit link $t_i=\{s_i,q_i\}$ from a batch of true issue-commit links $\mathcal{B}_\text{true}$, we sample another link $t_j=\{s_j,q_j\}$ from $\mathcal{B}_\text{true}$ and use the commit $q_j$ as the positive sample of $q_i$.
$t_i$ and $t_j$ belong to the same one-to-many link, i.e., $s_i$ and $s_j$ are identical.
If no such $q_j$ exists, $q_i$ is treated as the positive sample of itself.
We do not sample negative samples explicitly.
Instead, we treat other $\left| \mathcal{B}_{\text{true}} \right| - 2$ links within $\mathcal{B}_{\text{true}}$ (exclude all other links in the same one-to-many link as $t_i$ and $t_j$) as negative examples.

The representation $\mathbf{q}_i$ of a commit $q_i$ is obtained via: $\mathbf{q}_i = \mathbf{m}_i \oplus \mathbf{c}_i$, 
where $\oplus$ is the concatenation operation. $\mathbf{m}_i$ and $\mathbf{c}_i$ are representations of commit message and commit code generated by the student model, respectively. 
For contrastive instance discrimination, we adopt a non-parametric contrastive learning loss~\cite{WuXYL18} by using cosine similarity as similarity measure: 
\begin{equation}
	\mathcal{L}_{\text{cl}}(\mathcal{B}_\text{true})=-\sum_{t_i\in \mathcal{B}_\text{true} }\log \frac{e^{\operatorname{sim}\left(\mathbf{q}_i, \mathbf{q}_i^{+}\right) / \tau}}{\sum_{j\in \mathcal{N}(t_i)} e^{\operatorname{sim}\left(\mathbf{q}_i, \mathbf{q}_j\right) / \tau}},
\end{equation}
where $\mathcal{L}_{\text{cl}}(\mathcal{B}_\text{true})$ is the loss for true links $\mathcal{B}_\text{true}$ in a batch $\mathcal{B}$, $\mathcal{N}(t_i)$ is the in-batch negative samples of the link $t_i$, $\mathbf{q}_i^{+}$ denotes the representation of the positive sample of the commit $q_i$, and $\operatorname{sim}\left(\cdot\right)$ indicates the inner product. 
$\tau$ is a temperature hyperparameter that controls the concentration level of the distribution.
Summing up $\mathcal{L}_{\text{cl}}(\mathcal{B}_\text{true})$ for all batches, we have the complete contrastive learning loss $\mathcal{L}_{\text{cl}}$.

\subsection{Distinction between Commit Code (P3)}
\label{sec:mtl}

To eliminate the noise brought by loosely related and unrelated commit code, we define an auxiliary task, issue-code link prediction to enhance the understanding of \ours on the importance of each commit code. 
The issue-code link prediction task estimates the relevance between issues and commit code.

To be specific, each issue-code link $p_i=\left<s_i, c_i \right>$, where $c_i$ indicates the first $k_{\text{PL}}$ code tokens in the code change of a changed code file, is fed to the student model. 
$\mathbf{s}_i$ and $\mathbf{c}_i$ are representations of the issue text $s_i$ and the code $c_i$ in the link $p_i$, respectively.
They are the concatenation of their tokens' representations.
Similar to commit code, we preserve the first $k_{\text{NL}}$ tokens in each issue text.
Then, $\mathbf{s}_i$ and $\mathbf{c}_i$ are passed to a binary classifier which is a two-layer feedforward neural network (FFN): 
\begin{equation}
\begin{aligned}
\mathbf{s}_i'&=\text{AVG}(\mathbf{s}_i),\,\,\,\mathbf{c}_i'=\text{AVG}(\mathbf{c}_i)\\
\hat{\mathbf{y}}_{\left<s_i, c_i \right>} &=\text{f}_2\Big(\text{f}_1\big(\mathbf{s}_i'\oplus \mathbf{c}_i' \oplus \left| \mathbf{c}_i'-\mathbf{s}_i' \right|\big)\Big),
\end{aligned}
\end{equation}
where $\text{AVG}(\cdot)$ indicates average pooling, $\text{f}_1(\cdot)$ is a single-layer FFN with the Tanh activation, and $\text{f}_2(\cdot)$ is a single-layer FFN without activation. 
$\hat{\mathbf{y}}_{\left<s_i, c_i \right>}\in\mathbb{R}^2$ indicates the probabilities of the existence and the non-existence of the issue-code link $p_i$.

Based on the data we construct in Sec.~\ref{sec:auxiliary_data} for issue-code link prediction and the estimated probabilities $\mathbf{y}$ of the issue-code link, we train \ours with an auxiliary objective: 
\begin{equation}
	\mathcal{L}_{\text {aux}} = -\sum_{x\in X_{\text{aux}}} \sum_{\left<s,\,c\right> \in x} y_{\left<s,\,c\right>} \cdot \log \hat{\mathbf{y}}_{\left<s,\,c\right>},
\end{equation}
where $X_{\text{aux}}$ is the true issue-commit link set in the auxiliary dataset, $x$ is a true issue-commit link in $X_{\text{aux}}$, $s$ indicates the issue text in $x$, $c$ is the commit code in one changed file of a commit in $x$, and $y_{\left<s,\,c\right>}$ is the binary label for the relevance between $s$ and $c$.

We train the issue-code link prediction task together with the issue-commit link recovery task in a manner of multi-task learning~\cite{ZhangY22} (illustrate in Sec.~\ref{sec:opt}). 
Multi-task learning helps \ours learn from related tasks and it enhances the understanding of \ours on the importance of each commit code, avoiding the negative impact of loosely related and unrelated commit code.
Unlike FRLink~\cite{SunWY17} that directly filters commit code sharing few similar terms with issue text, our approach is learning based and does not filter any commit code.
This way, we do not need to manually set a similarity threshold for filtering commit code and avoid the case that the commit code is semantically related but it shares few similar terms with the issue text.
Moreover, loosely related and unrelated commit code is still leveraged in training: they help train \ours to better distinguish commit code.

\subsection{Putting All Together}
\label{sec:opt}

The main task of \ours (i.e., the issue-commit link recovery task) is optimized via the following objective used in existing works~\cite{RuanCPZ19}: 
\begin{equation}
	\label{eq:main_obj}
	\mathcal{L}_{\text{main}} = \sum_{x \in X}\sum_{\left<s,\,q\right> \in x}\left|{y}_{\left<s, q\right>} - sim(\mathbf{s}, \mathbf{q})\right|,
\end{equation}
where $X$ is the issue-commit link set, $x$ is a link, $s$ is the issue in $x$, $q$ is a commit in $x$, ${y}_{\left<s, q\right>}$ is the binary label (1 if $\left<s, q\right>$ is a true link, otherwise 0), and $sim(\cdot)$ is the cosine similarity.
$\mathbf{s}$ is the representation of $s$ output by the student model and $\mathbf{q}$ is the representation of $q$ obtained similar as in Sec.~\ref{sec:cl}.

\vspace{5pt}
\noindent\textbf{False Link Generation (P4):} $X$ includes both true links (extracted by tags) and false links. The main issue of existing false link generation methods~\cite{BirdBRB10,NguyenNNN12,SunWY17,RuanCPZ19} is that they require a time threshold (e.g., 7 days) to filter commits.
But it is hard to set a reasonable time threshold.
To overcome this problem, we opt to design a similarity based false link generation method. 
To be specific, for the issue $s$ in a known true link, \ours finds its least similar issue $s'$ by comparing cosine similarity between representations of their issue text. 
We connect the commit $q$, which is originally linked to $s$ in a true link, to $s'$ in order to construct a false link. 
If multiple possible $q$ exists (i.e., one-to-many issue-commit link), we pick the first commit scanned by \ours. 
The main overhead of the above method is caused by calculating all-pair cosine similarities and the time complexity is $O(\left|\mathcal{T}\right|^2)$ where $\left|\mathcal{T}\right|$ is the number of all issues.
To reduce the cost of generating false links, we only search irrelevant issues contained in the same batch of true links ($\mathcal{B}_\text{true}$) instead of all issues during training and the time complexity reduces to $O(\left|\mathcal{B}_\text{true}\right|^2)$.
\vspace{5pt}

As illustrated in Sec.~\ref{sec:mtl}, \ours is optimized in a manner of multi-task learning and the complete objective is as follows: 
\begin{equation}
	\label{eq:complete_loss}
	\mathcal{L}=\mathcal{L}_{\text{main}} + \lambda_{\text{cl}} \cdot \mathcal{L}_{\text{cl}} + \lambda_{\text{aux}} \cdot \mathcal{L}_{\text {aux}},
\end{equation}
where $\lambda_{\text{cl}}$ and $\lambda_{\text{aux}}$ are hyper-parameters for task weights. 
Through multi-task learning, \ours is fine-tuned to leverage useful information contained in multiple related tasks that may be difficult to learn in the main task to improve the recovery performance.

When making predictions, \ours calculates the cosine similarity between the representations $\mathbf{s}$ and $\mathbf{q}$ of the target issue $s$ and the target commit $q$.
$\mathbf{s}$ and $\mathbf{q}$ are obtained in a similar way as in Eq.~\ref{eq:main_obj}.
Following FRLink~\cite{SunWY17} and DeepLink~\cite{XieCYLHDZ19}, the similarity threshold is set to 0.5, which means that if the similarity is larger than $0.5$, \ours will predict $\left<s, q\right>$ as a true link.

\section{Experiments}
\label{sec:exp}

\begin{table*}[t]
	\centering
	\caption{Comparisons between different methods. Best performance on a project w.r.t. a metric is shown in bold.}
	\resizebox{1\textwidth}{!}{
		\begin{tabular}{c|cccccc|cccccc}
			\hline
			                              & \multicolumn{6}{c|}{\textbf{\ours}}                                                                                                                                                                  & \multicolumn{6}{c}{\textbf{T-BERT}}                                                                                                                                                      \\
			\textbf{Project}              & \textbf{P@1 (Hit@1)}           & {\color[HTML]{000000} \textbf{P@10}} & \textbf{Hit@10}               & \textbf{MRR}                  & \textbf{NDCG@1}               & \textbf{NDCG@10}              & \textbf{P@1 (Hit@1)}           & \textbf{P@10}                & \textbf{Hit@10}              & \textbf{MRR}                 & \textbf{NDCG@1}              & \textbf{NDCG@10}             \\ \hline
			{\color[HTML]{000000} Ambari} & \textbf{0.9490}               & \textbf{0.1231}                      & \textbf{0.9800}                 & \textbf{0.9622}               & \textbf{0.5988}               & \textbf{0.6130}               & 0.5524                        & 0.0775                       & 0.7748                       & 0.6321                       & 0.3485                       & 0.4405                       \\
			Calcite                       & \textbf{0.6555}               & \textbf{0.0920}                      & \textbf{0.8566}               & \textbf{0.7313}               & \textbf{0.4136}               & \textbf{0.4955}               & 0.5540                        & 0.0836                       & 0.8357                       & 0.6450                       & 0.3495                       & 0.4626                       \\
			Groovy                        & \textbf{0.6650}               & \textbf{0.1029}                      & \textbf{0.8310}               & \textbf{0.7323}               & \textbf{0.4196}               & \textbf{0.4919}               & 0.4655                        & 0.0718                       & 0.7175                       & 0.5482                       & 0.2937                       & 0.3937                       \\
			{\color[HTML]{000000} Ignite} & 0.3950                        & \textbf{0.0907}                      & \textbf{0.7210}               & 0.5054                        & 0.2492                        & 0.3814                        & \textbf{0.4304}               & 0.0709                       & 0.7087                       & \textbf{0.5273}              & \textbf{0.2716}              & \textbf{0.3843}              \\
			Isis                          & 0.2423                        & \textbf{0.0983}                      & \textbf{0.5184}               & 0.3397                        & 0.1529                        & 0.2622                        & 0.2394                        & 0.0443                       & 0.4304                       & 0.3049                       & 0.1510                       & 0.2221                       \\
			Netbeans                      & \textbf{0.3273}               & \textbf{0.0824}                      & \textbf{0.6727}               & \textbf{0.4567}               & \textbf{0.2065}               & \textbf{0.3519}               & 0.2794                        & 0.0507                       & 0.5074                       & 0.3614                       & 0.1763                       & 0.2638                       \\ \hline
			\textbf{Average}              & \textbf{0.5390}               & \textbf{0.0982}                      & \textbf{0.7633}               & \textbf{0.6213}               & \textbf{0.3401}               & \textbf{0.4327}               & 0.4202                        & 0.0665                       & 0.6624                       & 0.5032                       & 0.2651                       & 0.3612                       \\
			\textbf{Improve}              & -                             & -                                    & -                             & -                             & -                             & -                             & \textbf{($\uparrow$28.27\%)}  & \textbf{($\uparrow$47.67\%)} & \textbf{($\uparrow$15.23\%)} & \textbf{($\uparrow$23.47\%)} & \textbf{($\uparrow$28.29\%)} & \textbf{($\uparrow$19.80\%)} \\ \hline
			                              & \multicolumn{6}{c|}{\textbf{DeepLink}}                                                                                                                                                               & \multicolumn{6}{c}{\textbf{VSM}}                                                                                                                                                         \\
			\textbf{Project}              & \textbf{P@1 (Hit@1)}           & \textbf{P@10}                        & \textbf{Hit@10}               & \textbf{MRR}                  & \textbf{NDCG@1}               & \textbf{NDCG@10}              & \textbf{P@1 (Hit@1)}           & \textbf{P@10}                & \textbf{Hit@10}              & \textbf{MRR}                 & \textbf{NDCG@1}              & \textbf{NDCG@10}             \\ \hline
			Ambari                        & 0.0625                        & 0.0238                               & 0.2167                        & 0.1266                        & 0.0394                        & 0.0992                        & 0.4209                        & 0.0680                        & 0.6798                       & 0.5103                       & 0.2655                       & 0.3698                       \\
			Calcite                       & 0.0926                        & 0.0185                               & 0.1739                        & 0.1346                        & 0.0584                        & 0.0909                        & 0.2770                         & 0.0676                       & 0.7512                       & 0.4919                       & 0.1866                       & 0.3677                       \\
			Groovy                        & 0.0062                        & 0.0176                               & 0.1522                        & 0.0587                        & 0.0039                        & 0.0554                        & 0.4983                        & 0.0776                       & 0.7763                       & 0.5936                       & 0.3144                       & 0.4284                       \\
			Ignite                        & 0.0584                        & 0.0182                               & 0.1700                        & 0.1100                        & 0.0368                        & 0.0813                        & 0.1012                        & 0.0581                       & 0.5806                       & 0.3256                       & 0.1269                       & 0.2765                       \\
			Isis                          & \textbf{0.3291}               & 0.0538                               & 0.4960                        & \textbf{0.3994}               & \textbf{0.2076}               & \textbf{0.2787}               & 0.1191                        & 0.0471                       & 0.4705                       & 0.2264                       & 0.0751                       & 0.2065                       \\
			Netbeans                      & 0.0870                        & 0.0224                               & 0.1925                        & 0.1293                        & 0.0549                        & 0.0917                        & 0.0125                        & 0.0375                       & 0.3750                       & 0.2040                       & 0.0789                       & 0.1710                       \\ \hline
			\textbf{Average}              & 0.1060                        & 0.0257                               & 0.2336                        & 0.1598                        & 0.0668                        & 0.1162                        & 0.2382                        & 0.0593                       & 0.6056                       & 0.3920                       & 0.1746                       & 0.3033                       \\
			\textbf{Improve}              & \textbf{($\uparrow$408.65\%)} & \textbf{($\uparrow$281.85\%)}        & \textbf{($\uparrow$226.83\%)} & \textbf{($\uparrow$288.88\%)} & \textbf{($\uparrow$408.88\%)} & \textbf{($\uparrow$272.38\%)} & \textbf{($\uparrow$126.28\%)} & \textbf{($\uparrow$65.60\%)} & \textbf{($\uparrow$26.04\%)} & \textbf{($\uparrow$59.49\%)} & \textbf{($\uparrow$94.79\%)} & \textbf{($\uparrow$42.66\%)} \\ \hline
			\end{tabular}
	}
	\label{tab:overall}
\end{table*}

\subsection{Evaluation Settings} 
We choose four prevalent metrics Precision@$k$ (P@$k$), Normalized Discounted Cumulative Gain (NDCG@$k$), Mean Reciprocal Rank (MRR) and Hit Ratio (Hit@$k$) for evaluation.
In our evaluation, a query refers to a true link and its top-$k$ ranking list consists of $k$ ranked candidate commits.  
\begin{itemize} 
	\item {\textbf{Precision@$\mathbf{k}$}} indicates the percentage of relevant commits to the query in the top-K ranking list predicted by a model:  
	\begin{equation}
		\text{Precision}@k=\frac{1}{|\mathcal{Q}|} \sum_{i\in \mathcal{Q}}\frac{\operatorname{Rel}_i}{\operatorname{k}},
	\end{equation}
	where $\mathcal{Q}$ is the query set, $|\mathcal{Q}|$ is the number of queries, $\operatorname{Rel}_i$ represents the number of relevant commits that appear in the top-$k$ ranking list for the $i$-th query. 

	\item {\textbf{NDCG@$\mathbf{k}$}} evaluates the ranking position of relevant commits: 
	\begin{equation}
		\text{NDCG}@ k=\frac{1}{Z_k} \sum_{i=1}^k \frac{2^{r_i}-1}{\log _2(i+1)},
	\end{equation}
	where $Z_k$ is a normalizer which ensures that perfect ranking has a value of $1$, $r_i$ is the relevance of commit at position $i$. We use simple binary relevance: $r_i=1$ if the commit is actually linked to the target issue, and 0 otherwise. 
	
    \item {\textbf{MRR}} measures whether the relevant commits to an issue are placed in more prominent positions in the ranking list: 
	\begin{equation}
		\text{MRR}=\frac{1}{|\mathcal{Q}|} \sum_{|i=1|}^{|\mathcal{Q}|}\frac{1}{\operatorname{Rank}_i},
	\end{equation}
	where $\operatorname{Rank}_i$ refers to the ranking of the relevant commit.

	\item {\textbf{Hit@$\mathbf{k}$:}} refers to the probability that a relevant commit is in the predicted top-$k$ ranking list: 
	\begin{equation}
		\text{Hit}@k=\frac{1}{|\mathcal{Q}|} \sum_i^{|\mathcal{Q}|}\mathbb{I}\left(\operatorname{Rank}_{i} \leq k\right)
	\end{equation}
	where $\mathbb{I}(\cdot)$ is an indicator function. If the condition is true, it returns 1, otherwise 0.
	Hit@$1$ is identical to P@$1$.
	
\end{itemize}

\vspace{3pt}
For each project, true links were randomly divided by 6:2:2 for training, validation and test.  
To accelerate the evaluation, 
we randomly sampled 1,000 unique issues at most from each project's test set.
If the number of unique issues was less than 1,000, we selected all the unique issues. 
In each test set, true links $\mathcal{E}$ containing the selected issues were picked.
For each true link $t_i=\{s_i, q_i\}$ in $\mathcal{E}$, we randomly sampled 99 other true links in $\mathcal{E}$.
Let $t_j=\{s_j, q_j\}$ be one of the 99 sampled links.
$s_i$ and $s_j$ should not be identical, i.e., $t_i$ and $t_j$ do not belong to the same one-to-many issue-commit link.
Then, we connected $q_j$ to $s_i$ to construct a false link for $t_i$.
For each true link in $\mathcal{E}$, we constructed 99 false links.
In the experiments, we evaluate whether the model can rank a true link ahead of its corresponding 99 false links.

\subsection{Environment and Hyper-parameter Settings}
We implemented \ours using PyTorch.
The experiments were run on a machine with two Intel(R) Xeon(R) Silver 4214R CPU @ 2.40GHz, 256 GB main memory and one NVIDIA GeForce RTX 3090.

The default teacher model in \ours is CodeBERT~\cite{FengGTDFGS0LJZ20} and we used hyper-parameters provided by the authors.
We also explored other code pre-trained models in our experiments.
The student model had 2 layers, 1 head and 768 hidden dimensions.
By default, we set distillation channels as follows: the first layer of teacher was connected to the first layer of student, and the fifth layer of teacher was connected to the second layer of student. 
Results of other settings for distillation channels are reported in RQ4 of Sec.~\ref{subsec:RQ}.
$\lambda_{\text{cl}}$ and $\lambda_{\text{aux}}$ were set to 1 by default. 
The length of tokens in each natural language text (e.g., $k_{\text{NL}}$ in Sec.~\ref{sec:mtl}) and the length of tokens in each code data (e.g., $k_{\text{PL}}$ in Sec.~\ref{sec:mtl}) were set to 35 and 80, respectively. 
To train the model, we employed the Adam optimizer~\cite{KingmaB14}, and the batch size was set as 16. 
The initial learning rate was set to $4e^{-5}$ and was multiplied by 0.8 every 6 epochs. 
We set hyper-parameters for baselines as suggested by their authors.

\subsection{Baselines}
We compare \ours with three representative and publicly available issue-commit link recovery approaches:
\begin{itemize} 
	\item \textbf{T-BERT}\footnote{\url{https://github.com/jinfenglin/TraceBERT}}~\cite{LinLZ0C21} is the state-of-the-art pre-trained method. It is first pre-trained on the large-scale CodeSearchNet~\cite{abs-1909-09436} for the code search task~\cite{LiuXLGYG22}. Then, the pre-trained T-BERT is fine-tuned on the small issue-commit link data for the link recovery task. 
	 
	\item \textbf{DeepLink}\footnote{\url{https://github.com/ruanhang1993/DeepLink}}~\cite{RuanCPZ19} is a deep learning based method which adopts word embedding and RNN to learn the semantic representations of issues and commits for interlinking issues and commits. 

	\item \textbf{VSM}\footnote{We use \url{https://pypi.org/project/gensim} to implement VSM.}~\cite{LinLZ0C21,0001ZLWK22} is an information retrieval based method. It expresses code and textual data as bags of words which are further represented as word vectors. The relevance between an issue and a commit is calculated based on the similarity between their corresponding word vectors.
\end{itemize}
We set all the hyper-parameters for baselines as specified in the original papers.

\subsection{Experimental Results and Analysis}
\label{subsec:RQ}

Next, we report and analyze the results of our experiments in order to answer five research questions:

\vspace{5pt}
\noindent\textbf{RQ1: Does \ours outperform state-of-art baselines?}
\vspace{3pt}

We compare \ours with three baselines on the 6 Java projects in our dataset.
Tab.~\ref{tab:overall} reports the performance of all methods and the improvement percentage of \ours over baselines.
In Tab.~\ref{tab:overall}, the best performance on a project w.r.t. a metric is shown in bold.

Tab.~\ref{tab:overall} shows that, in a few cases, baselines outperforms EALink. However, they do not show robust performance as EALink. For example, T-BERT is 0.04 higher than EALink in one project Ignite on P@1 and  DeepLink outperforms EALink in the Isis project, while on average, \ours achieves 0.5390 on P@1 (Hit@1), 0.0982 on P@10, 0.7633 on Hit@10, 0.6213 on MRR, 0.3401 on NDCG@1 and 0.4327 on NDCG@10, respectively. 
The lowest and highest Hit@10 are 0.5184 for the Isis project and 0.9800 for the Ambari project, respectively. 
The high Hit@k indicates that the relevant commit has a high probability of being ranked in the top 10 by \ours. 
Compared to baselines, \ours improves the quality of issue-commit link recovery by 408.65\% at most, showing the effectiveness of \ours.

The pre-trained model T-BERT achieves the highest P@1, MRR, NDCG@1 and NDCG@10 on Ignite, while \ours achieves highest P@10 and Hit@10.
However, on the other five projects, \ours outperforms T-BERT on all metrics.
On average, \ours exceeds T-BERT by 15.23\%-47.67\%, showing that the superiority of \ours over the existing pre-training issue-commit recovery method.
Moreover, we can conclude that the adoption of knowledge distillation to compress the size of the pre-trained model does not sacrifice the accuracy of recovery.

Surprisingly, DeepLink shows worst performance on almost all metrics and its performance is even worse than traditional information retrieval based method VSM.
The exception is the Isis project, where DeepLink shows best P@1 (Hit@1), MRR, NDCG@1 and NDCG@10.
DeepLink adopts deep neural networks to better capture the semantics of code and textual data.
A possible reason for its subdued performance, in our opinion, is the design of the similarity module.
It computes the cosine similarities between issue title vector and commit message vector, issue description vector and commit message vector, and issue code vector and commit code vector, where the term ``vector'' refers to the representation.
Then, the maximum cosine similarity among the three calculated values is used to estimate the relevance between an issue and a commit.
However, the cross-modal similarity, i.e., the similarity between the issue and the changed code in commits is not considered.
Moreover, it chooses the maximum similarity value instead of considering all the three similarity values, which may cause information loss.

\vspace{5pt}
\noindent\textbf{RQ2: Does \ours reduce the training and inference overhead compared to existing pre-trained method? (P1)}
\vspace{3pt}

One design goal of \ours is to reduce the training and inference overhead of deep learning based link recovery method.
In Tab.~\ref{tab:time}, we report the training and test time of \ours, T-BERT and DeepLink on a moderate-size project Isis among all the 6 projects in our constructed dataset. As shown in Tab.~\ref{tab:time}, DeepLink needs 1,145 seconds to evaluate 1,000 recovery queries and T-BERT needs 44,353 seconds, which indicates that the pre-trained model is heavier and more time-consuming than the simple deep learning model. EALink uses only 126 seconds, which is several orders of magnitude lower than T-BERT. Hence, \ours is more scalable to large-scale projects in the real world.
In other 5 projects, similar overhead gaps can be observed. 
Observed from Tab.~\ref{tab:overall} and Tab.~\ref{tab:time}, we can conclude that the design (i.e., use a distilled, compacted student model) of \ours makes it possible to take the advantage of a large code pre-trained model while keeping low overhead: \ours shows similar or much higher performance compared to T-BERT, while it requires training and test time that is an order of magnitude shorter.
In other words, \ours is suitable for deployment in large-scale projects.

\begin{table}[t]
	\caption{Total training and test time for each model on Isis. Test time is recorded for 1,000 recovery queries.}
	\centering
	\begin{tabular}{cccc}
		\hline
		& \textbf{\ours} & \textbf{T-BERT} & \textbf{DeepLink} \\ \hline
		Train (hr) & 14h              & 138h                & 37h               \\
		Test (sec) & 126s             & 44,353s              & 1,145s             \\ \hline
	\end{tabular}
	\label{tab:time}
\end{table}

\vspace{5pt} 
\noindent\textbf{RQ3: Does each component in \ours contribute to its performance? (P2, P3, P4)}
\vspace{3pt}

\begin{table}[t]
	\caption{Results of the ablation study on Isis.}
	\centering
	\begin{tabular}{ccHHccH}
	\hline
	              & \textbf{P@1 (Hit@1)} & \textbf{Pr@10}  & \textbf{Hit@10} & \textbf{MRR}    & \textbf{NDCG@1} & \textbf{NDCG@10} \\ \hline
	\ours w/o aux & 0.2193               & \textbf{0.0988} & \textbf{0.5199} & 0.3172          & 0.1384          & 0.2547           \\
	\ours w/o cl  & 0.0521               & 0.0402          & 0.25            & 0.1246          & 0.0329          & 0.1066           \\
	\ours         & \textbf{0.2423}      & \textbf{0.0983} & 0.5184          & \textbf{0.3397} & \textbf{0.1529} & \textbf{0.2622}  \\ \hline
	\end{tabular}
	\label{tab:ablation}
\end{table}

\begin{table}[!t]
	\caption{Results using different false link generation on Isis.}
	\centering
	\begin{tabular}{ccHHccH}
	\hline
	       & \textbf{P@1 (Hit@1)} & \textbf{P@10}   & \textbf{Hit@10} & \textbf{MRR}    & \textbf{NDCG@1} & \textbf{NDCG@10} \\ \hline
	\oursf & 0.1843              & 0.0897          & 0.4312          & 0.2695          & 0.1163          & 0.2114           \\
	\ours  & \textbf{0.2423}     & \textbf{0.0983} & \textbf{0.5184} & \textbf{0.3397} & \textbf{0.1529} & \textbf{0.2622}  \\ \hline
	\end{tabular}
    \label{tab:false_link}
\end{table}

\begin{table}[!t]
	\centering
	\caption{Performance of using different channels on Isis.}
	\begin{tabular}{cc|cHHccH}
		\hline
		\textbf{t$_1$} & \textbf{t$_2$} & \textbf{P@1 (Hit@1)} & \textbf{P@10}   & \textbf{Hit@10} & \textbf{MRR}    & \textbf{NDCG@1} & \textbf{NDCG@10} \\ \hline
		1              & 5              & \textbf{0.2423}     & \textbf{0.0983} & \textbf{0.5184} & \textbf{0.3397} & \textbf{0.1529} & \textbf{0.2622}  \\
		3              & 7              & 0.2417              & 0.0946          & 0.4908          & 0.3092          & 0.1355          & 0.2417           \\
		4              & 9              & 0.2301              & 0.1035          & 0,5046          & 0.3216          & 0.1452          & 0.2508           \\
		6              & 12             & 0.2224              & 0.1006          & 0.5092          & 0.3235          & 0.1403          & 0.2534           \\ \hline
	\end{tabular}
	\label{tab:channel}
\end{table}

\begin{table}[!t]
	\caption{Performance of using different task weights on Isis.}
	\centering
	\begin{tabular}{cc|cHHccH}
		\hline
		$\lambda_{cl}$ & $\lambda_{aux}$ & \textbf{P@1 (Hit@1)} & \textbf{P@10}   & \textbf{Hit@10} & \textbf{MRR}    & \textbf{NDCG@1} & \textbf{NDCG@10} \\ \hline
		1              & 1               & \textbf{0.2423}     & 0.0983          & \textbf{0.5184} & \textbf{0.3397} & \textbf{0.1529} & \textbf{0.2622}  \\
		0.5            & 2.5             & 0.1595              & 0.0837          & 0.4417          & 0.2596          & 0.1006          & 0.2104           \\
		2              & 10              & 0.1187              & 0.0874          & 0,4678          & 0.2807          & 0.1190          & 0.2250           \\
		3              & 15              & 0.2193              & \textbf{0.0986} & 0.5169          & 0.32            & 0.1384          & 0.2538           \\
		4              & 20              & 0.2117              & 0.0966          & 0.5031          & 0.31            & 0.1335          & 0.2465           \\ \hline
	\end{tabular}
	\label{tab:loss_weight}
\end{table}

\begin{table}[!t]
	\caption{Performance of using different teachers on Isis.}
	\centering
	\begin{tabular}{c|cHHccH}
	\hline
	\textbf{Teacher} & \textbf{P@1 (Hit@1)} & \textbf{P@10}   & \textbf{Hit@10} & \textbf{MRR}    & \textbf{NDCG@1} & \textbf{NDCG@10} \\ \hline
	CodeBERT         & \textbf{0.2423}     & \textbf{0.0983} & \textbf{0.5184} & \textbf{0.3397} & \textbf{0.1529} & \textbf{0.2622}  \\
	GraphCodeBERT    & 0.2239              & 0.0943          & 0.4847          & 0.3171          & 0.1413          & 0.2429           \\
	UniXcoder        & 0.2362              & 0.0968          & 0.5114          & 0.3393          & 0.1490          & 0.2586           \\ \hline
	\end{tabular}
	\label{tab:teacher}
\end{table}

\begin{figure}[t]
    \centering
    \subfloat{{\includegraphics[width=0.47\columnwidth]{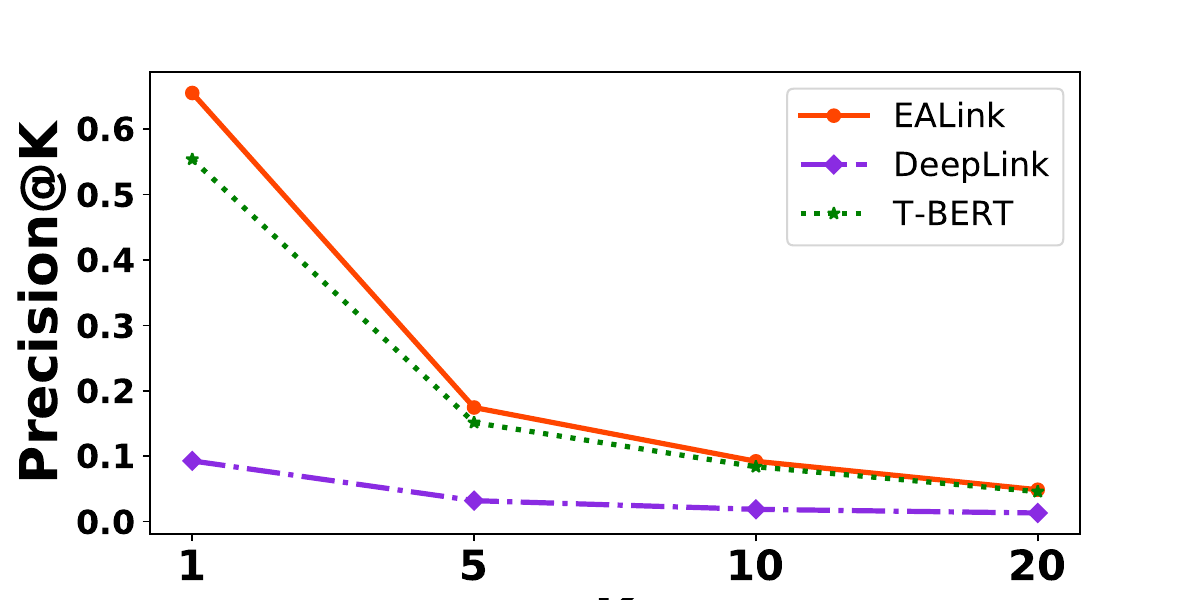} }}
    \subfloat{{\includegraphics[width=0.47\columnwidth]{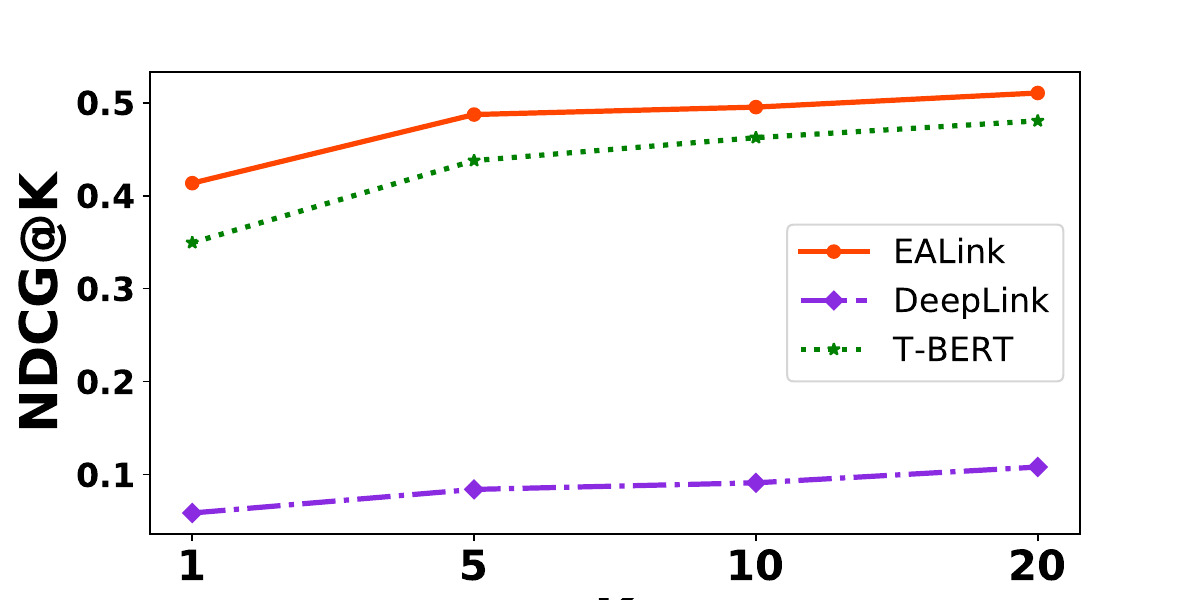} }}
    \caption{Impacts of $k$ on evaluation results. X axis shows $k$.}
    \label{fig:k}
\end{figure}

Firstly, we report the results of an ablation study in Tab.~\ref{tab:ablation}.
We take the Isis project as an example to illustrate and similar trends can be observed in other projects.
The ablation study allows us to evaluate the contribution of each components in \ours.
In Tab.~\ref{tab:ablation}, ``\ours w/o aux'' denotes the version that removes the issue-code prediction task proposed in Sec.~\ref{sec:mtl} and it treats all code changes in a commit equally.
``\ours w/o cl'' indicate the version that removes the inter-commit correlation modeling component illustrated in Sec.~\ref{sec:cl}.
From the results, we can observe that both ``\ours w/o aux'' and ``\ours w/o cl'' show worse performance compared to \ours: the MRR is reduced by 0.0225 and 0.2151 for Isis project, respectively.

The observation shows that, the designs of both the auxiliary task and the inter-commit correlation modeling enhance \ours.
In other words, addressing \text{P2} and \text{P3} discussed in Sec.~\ref{sec:motivation} helps improve the quality of generated links.
Comparing ``\ours w/o aux'' and ``\ours w/o cl'', we can see that the inter-commit correlation modeling component contributes more to the overall performance than the auxiliary task.

We further investigate the impact of using different false link generation methods and report the results on Isis in Tab.~\ref{tab:false_link}.
Similar conclusion can be drawn for other projects. 
In Tab.~\ref{tab:false_link}, \oursf indicates that we use the time interval based false link generation method used in previous link recovery methods~\cite{BirdBRB10,NguyenNNN12,SunWY17,RuanCPZ19}, which may generate conflicting false links as we discussed in Sec.~\ref{sec:motivation}.
We can see that \ours outperforms \oursf by a large margin since our proposed false link generation method produces high-quality false links and does not sample true links as false links.

In summary, based on the results reported in Tab.~\ref{tab:ablation} and Tab.~\ref{tab:false_link}, we can conclude that all components in \ours contribute to its high performance in the issue-commit link recovery task.

\vspace{5pt}
\noindent\textbf{RQ4: Do different settings affect \ours?}
\vspace{3pt}

We change several settings of \ours and investigate the impacts on \ours.
We use Isis to illustrate the result and similar trends can be observed in other projects.

Firstly, we investigate the impact of changing the distillation channels of the distillation component.
In Tab.~\ref{tab:channel}, we report the performance when varying the channels.
$t_1$ and $t_2$ indicate layers in the pre-trained teacher model that are chosen to establish the distillation channels.
The $t_1$-th layer is connected to the first layer in the student model and the $t_2$-th layer is linked to the second layer in the student model.
From Tab.~\ref{tab:channel}, we can see that changing the setting of channels affects the performance slightly and using the default setting yields the best result.

Secondly, we investigate the impacts of changing task weights $\lambda_{\text{cl}}$ and $\lambda_{\text{aux}}$ in the training objective shown in Eq.~\ref{eq:complete_loss}.
We report the results in Tab.~\ref{tab:loss_weight}.
When changing the values of $\lambda_{\text{cl}}$ and $\lambda_{\text{aux}}$, the performance of \ours is affected.
The best result is achieved using a balanced multi-task learning loss function ($\lambda_{\text{cl}}=1$, $\lambda_{\text{aux}}=1$).
Hence, we use $\lambda_{\text{cl}}=1$, $\lambda_{\text{aux}}=1$ in the default setting.

Thirdly, we change the pre-trained model used as the teacher in \ours and investigate the impact. 
Tab.~\ref{tab:teacher} demonstrates the performance of using three code pre-trained models CodeBERT~\cite{FengGTDFGS0LJZ20}, GraphCodeBert~\cite{GuoRLFT0ZDSFTDC21} and UniXcoder~\cite{GuoLDW0022} as the teacher in \ours on Isis.
We can see that using different teachers have an impact on the performance of \ours, but the performance differences are slight, showing that \ours is flexible to accommodate different code pre-trained models.
CodeBERT shows best results and we adopt it as the default teacher in \ours.

Finally, we further show the results when changing $k$. Fig.~\ref{fig:k} shows the performance when using different $k$ for evaluations. We can see that all the three methods exhibit similar trends as $k$ changes, showing that using different $k$ does not affect the performance rank of the three methods. In other words, using top-$k$ ranking metrics, we can draw consistent conclusions regardless of the value of $k$.

In summary, changing the settings of \ours affects its performance, but most changes do not incur significant impacts.

\section{Threats to Validity}

Two main threats may affect the validity of our study. 

The first threat is the types and the size of the training data. 
Our experiments are conducted on Java projects only.
We only train and evaluate recovery models on our constructed data from 6 Apache projects. 
The performance of \ours may vary when applying it to other programming languages or training it on more project data.
Moreover, the number of generated false links can be set to a larger number, while the number of true links is limited by the occurrence frequency of issue tags in commits.
To avoid the impact of imbalanced data on training \ours, we only generate one false link for each true link in \ours. 

The second threat is the reliability of true and false links.
Currently, true links in our dataset are established according to the crawled issue tags. 
But related commits may not contain issue tags, resulting in the missing of some true links in our data. 
We propose a false link generation mechanism used in \ours.
It connects the issue, which is least similar to the issue in a true link, to the commit in the corresponding true link for constructing a false link.
Since the used issue can be completely irrelevant to the commit in the true link, the mechanism may generate false links that are too easy for \ours to identify, bringing ineffective model training.


\section{Related Work}

In this section, we discuss several areas related to our work.

\subsection{Software Traceability Recovery}

Information retrieval (IR) methods are widely used in early approaches for traceability recovery.
For example, several works adopt Vector Space Model~\cite{HayesDS06}, Latent Semantic Indexing~\cite{LuciaFOT04,RempelMK13} and Latent Dirichlet Allocation~\cite{DekhtyarHSHD07,AsuncionAT10} to recover software traceability.
However, these methods heavily rely on feature engineering and represent software artifacts as bags of words, which cannot fully capture artifacts' semantics.
To solve this problem, machine learning (ML) methods are used to better capture artifacts' semantics. 
ENRL~\cite{FalessiPCC17} trains seven classifiers using golden standard traceability links to detect positive traceability links in the ranking lists generated by IR methods.
Similarly, TRAIL~\cite{MillsEH18} trains six classifiers to automatically verify ranked links generated by IR methods. 
SPLINT~\cite{0001ZLWK22} adopts semi-supervised learning to predict traceability links with unlabeled sets. 
To work with less training data, ALCATRAL~\cite{MillsEBKCH19} integrates active learning and supervised traceability link classifier. 
Comet~\cite{MoranPBMPSJ20} adopts probabilistic model to capture relationships between developer feedback and transitive (often implicit) relationships among groups of software artifacts.

\subsection{Issue-Commit Link Recovery}

Pioneering works on issue-commit link recovery~\cite{WuZKC11,NguyenNNN12,SchermannBPLG15} heavily rely on feature engineering and manual rules and are hard to generalize.
Traditional learning based methods adopt machine learning (e.g., classification) to automatically learn from features.
Representative approaches include but not limited to RCLinker~\cite{LeVLP15}, FRLink~\cite{SunWY17}, PULink~\cite{SunCWB17}, Hybrid-Linker~\cite{MazraeIH21} and the work of Rath et al.~\cite{0002RGCM18}.
Although they overcome the drawbacks of feature based and rule based approaches to some extent, they suffer from the deficiency of shallow learning (weak express power) and lack of training data.

More recent works deploy deep learning to improve issue-commit link recovery.
DeepLink~\cite{RuanCPZ19} adopts word embedding and RNN to learn the semantic representations of issues and commits.
TraceNN~\cite{0004CC17} uses similar techniques as DeepLink, but it is designed for recovering traceability links between requirements and design documents instead of issue-commit links. 
KG-DeepLink~\cite{XieCYLHDZ19} combines RNN and SVM for predicting links based on a code knowledge graph constructed from ASTs.  
Beside, pre-trained models like BERT~\cite{DevlinCLT19} also shed some light on overcoming the lack of issue and commit training data.
L{\"{u}}ders et al.~\cite{LudersPM22} adopt BERT to predict the types of links (e.g., duplicate and clone) in issue trackers.
T-BERT~\cite{LinLZ0C21} is first pre-trained on CodeSearchNet~\cite{abs-1909-09436}. Then, it is fine-tuned on small issue-commit link data for the link recovery task. 
Although deep learning based methods show promising results, they do not fully address the problems discussed in Sec.~\ref{sec:motivation}.

\subsection{Pre-training for Code Representation Learning}

There are various code-related tasks (e.g., code completion~\cite{WangL21a}, code refactoring~\cite{LiuWWXWLJ23} and code summarization~\cite{LinOZCLW21}) that require code representation learning.
The success of BERT~\cite{DevlinCLT19} has inspired the research on code pre-trained models that significantly benefit code-related tasks.
CodeBERT~\cite{FengGTDFGS0LJZ20} is a bimodal (NL and PL) pre-trained model. 
GraphCodeBERT~\cite{GuoRLFT0ZDSFTDC21} uses data flow to learn more comprehensive representations.
UniXcoder~\cite{GuoLDW0022} leverages cross-modal contents and mask attention matrices with prefix adapters to improve code pre-training.
Unlike previous code pre-trained models that only focus on the encoder, Mastropaolo et al.~\cite{MastropaoloSCNP21} and Niu et al.~\cite{NiuL0GH022} explore the seq2seq architecture in code pre-training.


\section{Conclusion}

Issue-commit links play a vital role in various software development and maintenance tasks. 
However, they are commonly deficient in software development and maintenance.
Therefore, automatic issue-commit link recovery methods, which can reduce the cost of manual labeling and assist with various software engineering tasks, have attracted significant attention.
In this paper, we point out the problems of existing issue-commit link recovery methods and propose \ours for efficiently and accurately recovering issue-commit links.
\ours requires much fewer model parameters but shows better recovery results compared to existing recovery models.
In the future, we will explore using other software artifacts (e.g., requirements, designs and test cases) to further improve \ours.
We will also deploy other model compression techniques to further reduce the model size of \ours and expedite the recovering process.

\section*{Acknowledgment}
This work was partially supported by National Key R\&D Program of China (No. 2022ZD0118201), National Natural Science Foundation of China (No. 62002303, 42171456),  Natural Science Foundation of Fujian Province of China (No. 2020J05001) and CCF-Tencent Open Fund.

\bibliographystyle{IEEEtran}
\balance
\bibliography{ref}

\end{document}